\documentclass[12pt]{article}
\usepackage{amssymb}
\usepackage{epsfig}
\newcommand     \ba             {\begin{eqnarray}}
\newcommand     \be             {\begin{equation}}

\newcommand     \ea             {\end{eqnarray}}
\newcommand     \ee             {\end{equation}}

\newcommand     \epem           {\ifmmode{e^+e^-}\else{$e^+e^-$}\fi}

\newcommand     \lambdamsb     {\ifmmode
          \Lambda_5^{\rm \scriptscriptstyle \overline{MS}} \else
         $\Lambda_5^{\rm \scriptscriptstyle \overline{MS}}$ \fi}
\newcommand     \Lambdamsb      \lambdamsb
\newcommand     \LambdaQCD     {\ifmmode
          \Lambda_{\rm \scriptscriptstyle QCD} \else
         $\Lambda_{\rm \scriptscriptstyle QCD}$ \fi}

\newcommand     \MSB            {\ifmmode {\overline{\rm MS}} \else
                                 $\overline{\rm MS}$  \fi}

\newcommand     \ptmin     {\ifmmode p_{\scriptscriptstyle T}^{\sss min} \else
                           $p_{\scriptscriptstyle T}^{\sss min}$ \fi}

  \newcommand{\ccaption}[2]{
    \begin{center}
    \parbox{0.85\textwidth}{
      \caption[#1]{\small{\it{#2}}}
      }
    \end{center}
    }
\newcommand\sss{\scriptscriptstyle\rm}

\newcommand\muf{\mu_{\sss F}}
\newcommand\mur{\mu_{\sss R}}

\newcommand\as{\alpha_{\sss S}}

\newcommand\mq{m_{\sss \rm Q}}

\def \pt   {p_{\scriptscriptstyle T}}

\def \to   {\mbox{$\rightarrow$}}
\def    \mb             {\mbox{$m_b$}}

\def\blue{}
\def\red{}

\newcommand\equationcolor{}

\newcommand\quotecolor{}

\newcommand\sssrm{\scriptscriptstyle\rm}

\begin{document}
\topskip 2cm 
\begin{titlepage}
{\flushright{
        \begin{minipage}{4cm}
        IFUM 638/FT \hfill \\
        hep-ph/9811468\hfill \\
        \end{minipage}        }

}
\vskip 3cm
\begin{center}
{\large\bf HEAVY FLAVOUR PRODUCTION\footnote{
Talk presented at the 12$^{\rm th}$ Rencontre de Physique de la Valle d'Aoste,
La Thuile, March 1998.}} \\
\vspace{2.5cm}
{\large Paolo Nason}
 \\
\vspace{.5cm}
{\sl CERN\footnote{
From November 1$^{\rm st}$, INFN, Milan, Italy}}\\
\vspace{2.5cm}
\vfil
\begin{abstract}
I review recent developements in the theory of heavy flavour production.
In particular, I dicuss the next-to-leading resummation of soft
gluon effects in heavy flavour hadroproduction, and the next-to-leading
resummation of collinear radiation
in the production of heavy flavour at
large transverse momenta.
\end{abstract}
\end{center}
\end{titlepage}
\section{Introduction}
In this talk I will review recent progress in heavy flavour production.
Current interest in this field is expecially motivated by
the need to predict cross sections and estimating backgrounds
for collider physics, and by its potential for QCD tests and
for constraining the parton densities. The availability of relatively
light (charm), moderately heavy (bottom), and very heavy (top) quarks
opens the possibility of studying the impact of low scale, non perturbative
effects in the production mechanism.

Heavy quark production cross sections in hadronic collisions are known
to order $\as^3$
for the total cross section \cite{NDE1} \cite{Beenakker1},
for the 1-particle inclusive distributions \cite{NDE2} \cite{Beenakker2},
and for the 2-particle inclusive distributions \cite{MNR}.
A recent review has appeared in
\cite{Buras2}, where a comprehensive bibliography can also be found.

Recently, calculations of heavy flavour
production in $\epem$ annihilation at order $\as^2$
have been completed \cite{Rodrigo} \cite{Bernreuther} \cite{Oleari}.
Their most direct applications are in the field of jet
studies. They are sometimes referred to as $b$ mass measurements
from jets \cite{RunningRodrigo} \cite{DELPHIbrun},
or as studies of the flavour independence of the strong charge
\cite{Chrisman}.

Other interesting applications are the study of
correlations in $b\bar{b}$ momenta \cite{bbcorr},
which is an important systematics in the study of $R_{\sssrm b}$,
and fragmentation function studies \cite{Frag97}.

It is quite clear that other important applications
will arise with future $\epem$ colliders above the $t\bar{t}$
threshold.

In this talk I will discuss in more detail problems that go beyond the
fixed order perturbation theory.
In particular, I will focus on a recent calculation of
soft gluon resummation effects at next-to-leading order
\cite{Bonciani}, and on the
high transverse momentum heavy flavour production \cite{CacciariGrecoNason}.

\section{Limitations of Fixed Order Results}
In special kinematic regions, large logarithms in the coefficients
of the perturbative expansion spoil its convergence. In particular, for
\begin{itemize}
\item
  {\red Total cross-sections:}
  \begin{itemize}
  \item
    At very high energy, one expects terms of order
    {\red $\mbox{Born}\times [\as\log S/\mq^2]^n$}
    to arise at all orders in perturbation theory. This problem, related
    to the small-x problem in DIS physics, is relevent, for example, for bottom
    production at the Tevatron and at the LHC.
  \item
    When we approach the threshold region for the production of the
    heavy flavour pair, terms of order
    {\red $\mbox{Born}\times [\as\log^2 (1-4\mq^2/S)]^n$} arise.
    These terms are relevant for the production of top at the Tevatron,
    for the production of $b$ at HERAb, and for the production
    of charm at relatively low CM energy.
  \end{itemize}
\item
  {\red Differential distributions:}
  \begin{itemize}
  \item
    For the production of heavy flavour at high transverse momenta
    one expects terms of order
    {\red $\mbox{Born}\times [\as\log \pt/\mq^2]^n$}
    to arise at all orders in perturbation theory. This problem
    is relevant for the production of charm and bottom at high
    transverse momentum at the Tevatron, but also, perhaps,
    for the production of very high transverse momentum top quarks
    at the LHC.
  \item
    For the production of a heavy flavour pair, when the transverse momentum
    of the pair is small, one expects terms
    of order
    {\red $\mbox{Born}\times [\as\log^2 \pt/\mq]^n$.}
    The resummation of these terms is analogous to the problem of
    computing the transverse momentum of the $W$ in hadronic collisions.
  \end{itemize}
\end{itemize}
Most of these problems have received some attention in the literature.
In this talk I will discuss the resummation of logarithms
arising from the threshold region, and the $\pt$ spectrum of
a heavy quark, since some progress has been made recently on these two topics.
\section{Soft Gluon Resummation For Total Cross Sections}
QCD favours soft gluon radiation.
\begin{figure}[htbp]
  \begin{center}
    \epsfig{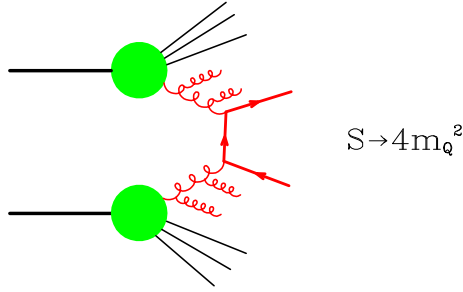}
    \caption{Soft gluon emission in heavy flavour production near threshold.}
    \label{fig:hvqprod1}
  \end{center}
\end{figure}
When radiation is restricted because of phase-space constraints,
the cross section is reduced ({\red Sudakov suppression}).
Suppression may become enhancement, depending upon how big a fraction
of the gluon radiation effects are incorporated already in the structure
functions, and must therefore be subtracted.
\subsection{How the logarithms are organized}
The organization of the large logarithms generated by soft gluon
emissions is as follows
\begin{equation}
\sigma = \sigma_0
 \times \exp\Bigg[ \log \xi\;\; g_1(\as\log\xi)+ g_2(\as\log\xi)
+{\cal O}(\as (\as\log\xi)^k) \Bigg]
\end{equation}
where 
$\xi = m_{\rm th}/\mq$, and $m_{\rm th}$ is a mass parameter
related to the typical distance from threshold of the
production process. For example, one may think of the $m_{\rm th}$ as the
average value of {\red $(\hat{s}-4\mq^2)/2\mq$}.
The functions $g_1$ and $g_2$ have a power expansion in their argument:
\begin{equation}
g_{1/2}(\as\log\xi)=g_{1/2}^{(1)}\;\as\log\xi + g_{1/2}^{(2)}\;(\as\log\xi)^2
 + \ldots\;.
\end{equation}
Leading logarithmic (LL) resummation includes only the term $g_1$ in the
exponent, while next-to-leading (NLL) resummation
includes also the term $g_2$.

In heavy quark production, the Born cross-section $\sigma_0$
is of order $\as^2$, and the resummation formula has the expansion
\begin{equation}
  \sigma = \sigma_0\times\left(1+g_1^{(1)}\as\log^2\xi+\ldots \right)
\end{equation}
Since the Born and ${\cal O}(\as^3)$ terms are known exactly,
one can replace them with the exact result. This procedure leads to
a next-to-leading order (NLO), LL resummed cross section.
Several groups have developed and
applied the relevant formulae \cite{LaenenSmithVanNeerven}
\cite{BergerContopanagos} \cite{CataniManganoNasonTrentadue},
and dangerous pitfalls to avoid when implementing the resummation
have been discussed in \cite{CataniManganoNasonTrentadue}.

It turns out that, with the addition of the LL resummation,
no improvement in the scale dependence can be expected.
In fact, for example, the scale in the value of $\as$
in the resummation formula is not specified exactly,
since its scale variation is of NLL order:
\begin{equation}
  \frac{d}{d\log\mu^2} \log\xi\; g_1(\as\log\xi)=
   -b_0\as^2\log^2\xi\; g_1^\prime(\as\log\xi)
\end{equation}
(where I used the evolution equation $d\as/d\log\mu^2=-b_0\as^2$).
\subsection{NLO NLL resummed result}
If NLL terms are included, one can expect a reduction in the scale dependence.
Formulae for the NLL soft gluon resummation have been obtained by
\begin{itemize}
\item 
{\quotecolor  Kidonakis, Sterman:} for the case of a heavy quark pair
produced at fixed invariant mass \cite{kidon}.
\item
{\quotecolor  Bonciani, Catani:} directly for the case of the
total cross section \cite{BoncianiThesis}. This avoids the need to perform
the angular integration numerically.
\end{itemize}
Phenomenological prediction have been obtained in
\cite{Bonciani}.

\begin{figure}[htbp]
  \begin{center}
    \epsfig{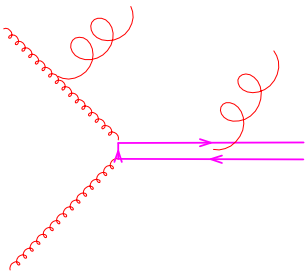}
    \caption{Intuitive graphical illustration of soft gluon
           radiation near threshold. Long wavelength gluons
           see the heavy quark pair as a single object.}
    \label{fig:hvqprod2}
  \end{center}
\end{figure}
When the $Q\bar{Q}$ pair is emitted near threshold, the quark and
antiquark are at rest with respect to each other, and thus behave as a
single particle as far as soft gluon emission is concerned
(see fig.~\ref{fig:hvqprod2}).  A
$Q\bar{Q}$ pair can be in a {\red color singlet} state, that is to
say, in a globally colorless state. In this case it does not emit
soft gluons: resummation is the same as for the Drell-Yan process.
Or, it can be in a {\red color octet} state. In this case it can emit.
Thus the resummation is performed by separating out the singlet and
octet part of the Born cross section, and applying different
soft-gluon resummation factors to them.

\begin{figure}[htbp]
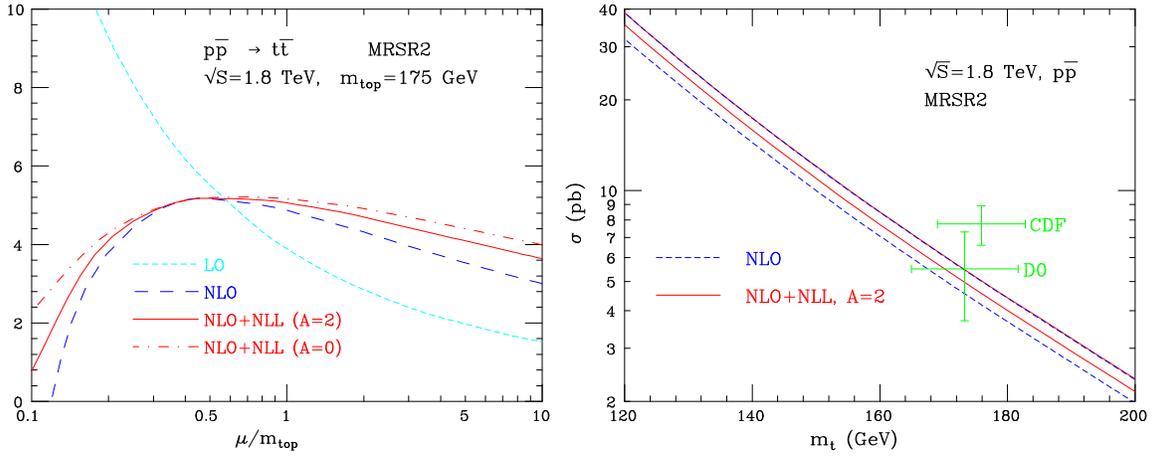

  \begin{center}
    \leavevmode
    \epsfig{file=tscale.eps,height=6cm}
    \epsfig{file=top18.eps,height=6cm}
    \caption{Resumamtion results for the top cross section}
    \label{fig:topres}
\end{center}
\end{figure}

\begin{figure}[htbp]
  \begin{center}
    \leavevmode
    \epsfig{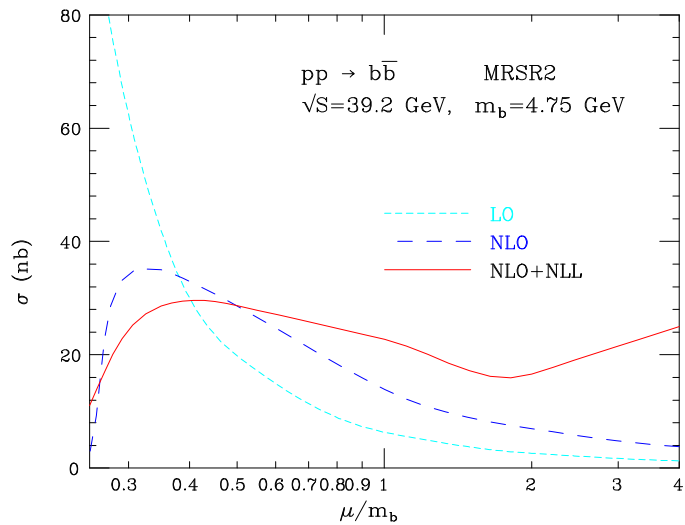}
    \caption{Scale dependence in bottom production at HERAb.}
    \label{fig:herab}
  \end{center}
\end{figure}

The scale dependence is considerably {\red reduced} after the inclusion of NLL
resummation. This is illustrated in fig.~\ref{fig:topres} and in
table~\ref{tab:topxs} for top production at the Tevatron.
\newcommand\mt{m_{\rm top}}
\begin{table}
\begin{center}
\begin{tabular}{|c|cc|cc|} \hline           
& \multicolumn{2}{c|}{$p\bar p$ at $\sqrt{S}=1.8$~TeV }
& \multicolumn{2}{c|}{$p p$ at $\sqrt{S}=14$~TeV }\\
\cline{2-5}
 $\mur=\muf$   & {\blue NLO} & {\red NLO+NLL} & {\blue NLO} & {\red NLO+NLL}\\
 \hline\hline
$\mt/2$     &  {\blue 5.17}  &{\red 5.19}   &{\blue 893}  &{\red 885}  
\\ \hline                                             
$\mt$   &  {\blue 4.87}  &{\red 5.06}   &{\blue 803} &{\red 833}   
\\ \hline                                             
$2\mt$ &  {\blue 4.31}  &{\red 4.70}   &{\blue 714} &{\red 794}
\\ \hline                                          
\end{tabular}
\ccaption{}{\label{tab:topxs} Total $t\bar t$ cross-sections ($m_t=175$~GeV)
at the Tevatron and LHC, in pb. PDF set MRSR2.} 
\end{center}      
\end{table}
The new uncertainty band at the Tevatron, including resummation effects,
is contained in the NLO one, confirming the fact that soft gluon
effects are not large for top production at the Tevatron.
Analogous results for bottom production at HERAb are shown in
fig.~\ref{fig:herab} and in table~\ref{tab:herab1}.
Also in this case, NLL resummation leads to an improvement in the
precision of the theoretical results.
\begin{table}
\begin{center}
\begin{tabular}{|c|cc|cc|} \hline
& \multicolumn{2}{c|}{$m_b$=4.5 GeV} 
& \multicolumn{2}{c|}{$m_b$=5.0 GeV} \\                             
\cline{2-5}                   
 $\mur=\muf$   & {\blue NLO} & {\red NLO+NLL} & {\blue NLO} & {\red NLO+NLL} \\
 \hline\hline
$\mb/2$ &   {\blue 27.4} & {\red 27.3} & {\blue 11.2}  & {\red 11.5} \\     
\hline                                             
$\mb$   &   {\blue 15.7}  & {\red 22.8} & {\blue 6.26}  & {\red 9.75} \\
\hline                                              
$2\mb$ &    {\blue 8.74}  & {\red 18.0} & {\blue 3.43}  & {\red 7.63} \\
\hline                                              
\end{tabular}
\ccaption{}{\label{tab:herab1} Total $b\bar b$ cross-sections (in nb)
at HERAB ($pp$ at $\sqrt{S}=39.2$~GeV), as a function of the the $b$ mass 
$m_b$. PDF set MRSR1.}
\end{center}      
\end{table}
\section{Resummation of large logarithms of the transverse momentum}
\begin{figure}[htbp]
  \begin{center}
    \leavevmode
    \epsfig{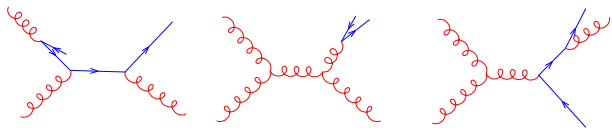}
    \caption{Enhanced collinear regions in heavy flavour production
     at large transverse momenta}
    \label{fig:highptenhanced}
  \end{center}
\end{figure}

It is known that at order $\as^3$ there are graphs
(such as the ones shown in fig.~\ref{fig:highptenhanced}) that
give cross section contributions enhanced by a factor $\log \pt/\mq$.
It can be shown that at higher and higher order, the structure of these
logarithms is as follows
\begin{eqnarray}
  \frac{d \sigma}{d\pt}&=& \frac{d \sigma_0}{d\pt}\times
\Bigg(1+\sum_{i=1}^\infty c_i^{(1)}(\as\log \pt/\mq)^i
\nonumber \\  &&
+\sum_{i=1}^\infty c_i^{(2)}\,\as\,(\as\log \pt/\mq)^i +\ldots \Bigg)\,,
\end{eqnarray}
where one usually refers to the terms up to to first sum as
leading-logarithmic terms (LL), and up to the second sum as
next-to-leading logarithmic (NLL), and so on.
The formalism for the resummation of these logarithms is the same as
the formalism for the inclusive production of hadrons at large transverse
momenta, based upon the factorization theorem formula
{\equationcolor
\begin{displaymath}
  \frac{d \sigma_h}{d\pt^2}=\sum_{ijk}
\int dx_1\,dx_2\,\frac{dz}{z^2} f_i(x_1)\,f_j(x_2)\,D_k^{(h)}(z)
\frac{d \hat\sigma_{ij\to k}}{d\hat\pt^2}\;.
\end{displaymath}}
where the structure functions {\red $f$}, the fragmentation functions {\red
  $D$}, and the partonic cross section {\red $\hat\sigma$}, all depend
upon a factorization and renormalization scale {\red $\mu$}. In the
case of a light hadron, {\red $i$, $j$} and {\red $k$} can be any
parton, and the structure functions and fragmentation functions must
be measured.

The same formalism applies when one looks at the production of a
heavy quark at large transverse momentum, the only differences being that
\begin{itemize}
\item
The fragmentation function {\red $D^{(h)}_i$} is replaced by
the heavy quark fragmentation function {\red $D^{(Q)}_i$}. All partons
{\red $i$, $j$ and $k$} may also be the heavy quark or antiquark.
\item
The fragmentation function {\red $D_i^{(Q)}$}
 for the heavy quark can be computed \cite{MeleNason}.
\item
The parton density for finding a heavy quark in a hadron can be computed
\cite{CollinsTung}.
\end{itemize}
 
The formalism for the production of high momentum heavy quarks in
$\epem$ annihilation has been developed in \cite{MeleNason}.
The formalism for the production of hadrons at large transverse
momenta in hadronic collisions have been developed at NLO in \cite{Aversa89},
and the formalism for the production of heavy quark at large transverse
momentum in hadronic collisions has been put together
in \cite{Cacciari94}.

Many applications to LEP physics, hadron collider physics, and HERA physics
have appeared.

The problem that has remained unsolved until now is the following:
The fragmentation function formalism neglects mass effects.
It is never clear at what value of $\pt$ mass effects are truly negligible.

\begin{figure}[htbp]
  \begin{center}
    \leavevmode
    \epsfig{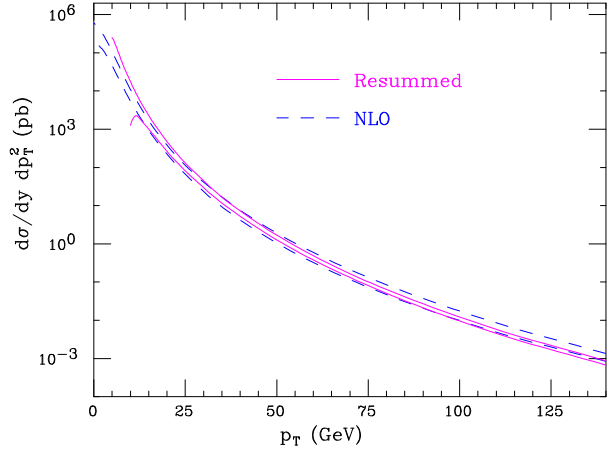}
    \caption{Comparison between the resummed and fixed order
        bottom production cross section at large transverse momenta,
        at the Tevatron.}
    \label{fig:resfo}
  \end{center}
\end{figure}  
It has been known from quite some time that the fixed order (NLO)
and the resummed calculation of {\quotecolor Cacciari and Greco}
match nicely at values of $\pt$ above 10 GeV (see fig.~\ref{fig:resfo}).
This would suggest that
mass corrections are in fact small. As we will see, instead, the 
matching at 10 GeV is quite accidental, and mass corrections
are large.
\section{The matched calculation}
The procedure can be schematically defined as follows.
The NLO calculation gives the terms
\begin{displaymath}
  \frac{d\sigma}{d\pt^2}=A(m)\as^2+B(m)\as^3
  +{\cal O}(\as^4)
\end{displaymath}
The explicit dependence upon $\hat{s}$, $\pt$ and $\mu$ is not indicated,
and $\as=\as(\mu)$.
The resummed calculation gives the terms
\begin{eqnarray}
  \frac{d\sigma}{d\pt^2}&=&
        \as^2\sum_{i=0}^\infty a_i (\as\log \mu/m)^i
   +    \as^3\sum_{i=0}^\infty b_i (\as\log \mu/m)^i 
\nonumber \\
&+& {\cal O}(\as^4(\as\log \mu/m)^i) + {\cal O}(\as^2 (m/\pt)^q)\;.
\end{eqnarray}
If $\mu \approx \pt \approx \hat{s}$, the coefficients $a_i$ and $b_i$
are not large (i.e., they contain no large logarithms).
The only large logarithms are explicitly indicated.
Our matched calculation should give
{\equationcolor
\begin{eqnarray}
\frac{d\sigma}{d\pt^2}&=& A(m)\as^2+B(m)\as^3+
        \as^2\sum_{i=2}^\infty a_i (\as\log \mu/m)^i
   +    \as^3\sum_{i=1}^\infty b_i (\as\log \mu/m)^i 
\nonumber \\&+&
 {\cal O}(\as^4(\as\log \mu/m)^i) + {\cal O}(\as^4 (m/\pt)^q) \nonumber
\end{eqnarray}}

  This looks simple, but it is in fact not so simple, since the
the various terms in the resummed calculation are buried inside
the structure functions and the fragmentation functions for the heavy quark.

We thus adopted the following approach: we computed the relevant combinations
of $a_0$, $a_1$ and $b_1$, from the relations
{\equationcolor
\begin{eqnarray}
A(m) &=& a_0 + {\cal O}( (m/\pt)^q) \nonumber \\
B(m) &=& a_1 \log \mu/m + b_0 +  {\cal O}( (m/\pt)^q)\nonumber
\end{eqnarray}}
In other words, with a mathematical procedure, we find the massless
limit of the massive calculation. Notice that this is not a real
limit, since there are logarithms of the mass that we must keep.
In the following
{\red
\begin{itemize}
\item FO: fixed order calculation
\item FOM0: massless limit (in the above sense) of the fixed order calculation
\item RS: resummed calculation
\item FONLL: FO-FOM0+RS, the result we seek
\end{itemize}}

\begin{figure}[htbp]
  \begin{center}
    \leavevmode
    \epsfig{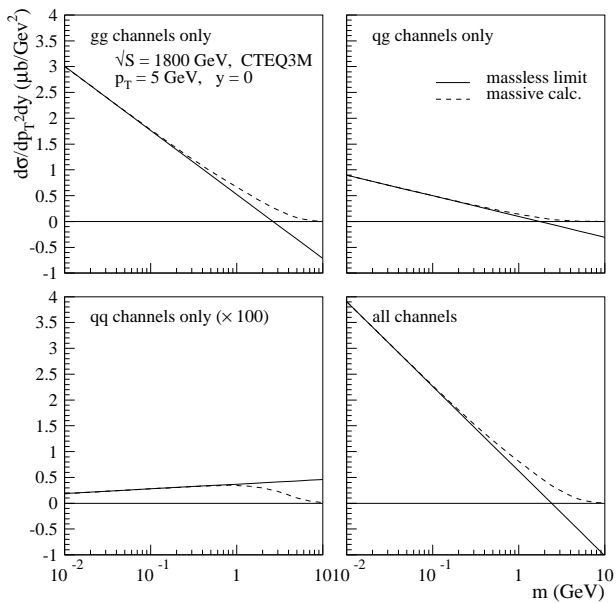}
    \caption{Massless limit of the fixed order calculation for
      the various incoming partons contributions. The straight lines
      represent the FOM0 approach.}
    \label{fig:masslesslim}
  \end{center}
\end{figure}

\begin{figure}[htbp]
  \begin{center}
    \leavevmode
    \epsfig{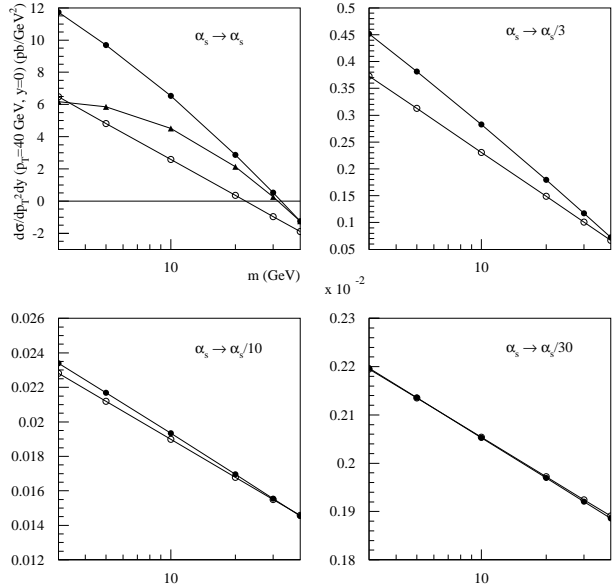}
    \caption{Cross section as a function of the mass at fixed $\pt$,
         for various values of $\as$. The empty circles represent
         the FOM0 calculation, the triangles represent the exact RS
         calculation, and the full circles represent an appropriate
         approximation to the RS calculation
         (see ref~\cite{CacciariGrecoNason} for details).}
    \label{fig:small_as}
  \end{center}
\end{figure}

  The FOM0 result is a linear function of $\log m$, everything else
being kept fixed.
As can be seen from fig.~\ref{fig:masslesslim},
mass effects are large: for $m=2$GeV and $\pt=5$ we still see substantial
(100\%) effects.
Another consistency check is shown in fig.~\ref{fig:small_as}, which shows
that the RS calculation tends to coincide with the FOM0 calculation
for small $\as$.

\begin{figure}[htbp]
  \begin{center}
    \epsfig{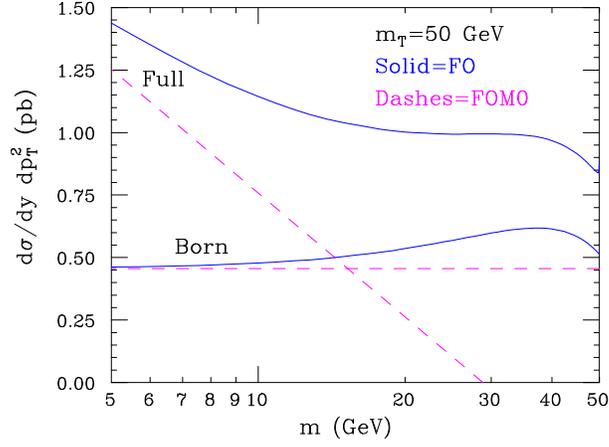}
    \caption{Differential cross section at fixed transverse mass,
       as a function of the mass. It is clear from the figure
       that the massless approximation is very bad for
       $\pt/m \lesssim 5$.}
    \label{fig:e_plot}
  \end{center}
\end{figure}

\begin{figure}[htbp]
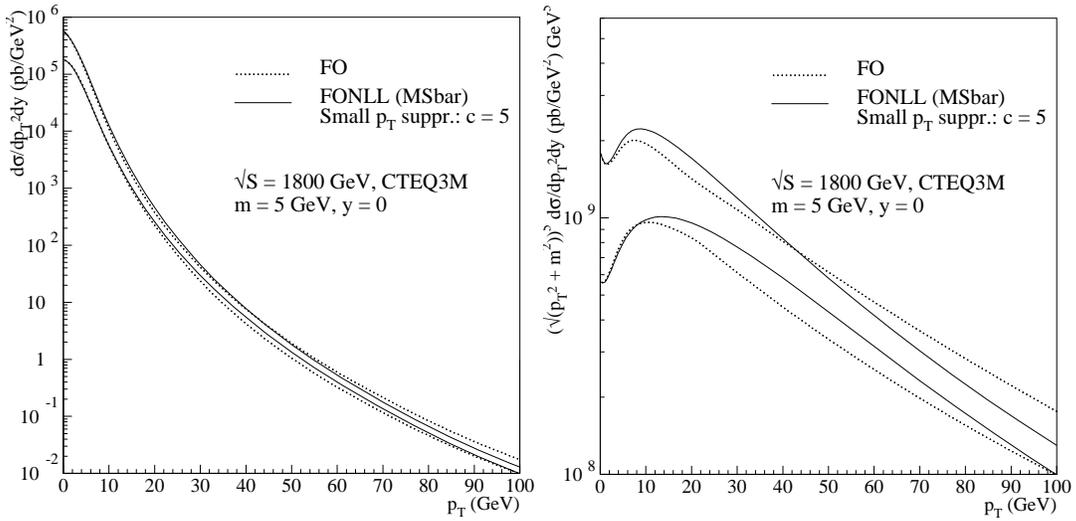

  \begin{center}
    \epsfig{file=band-ms-c5.seps,width=7cm}
    \epsfig{file=band-ms-c5-mt5.seps,width=7cm}
    \caption{Comparison between the matched calculation and the fixed order
         one. A slight anhancement for intermediate values of the transverse
         momentum is present. At very large transverse momenta, a sensible
         reduction of scale dependence is found.}           
    \label{fig:band-ms-c5}
  \end{center}
\end{figure}

\begin{figure}[htbp]
  \begin{center}
    \epsfig{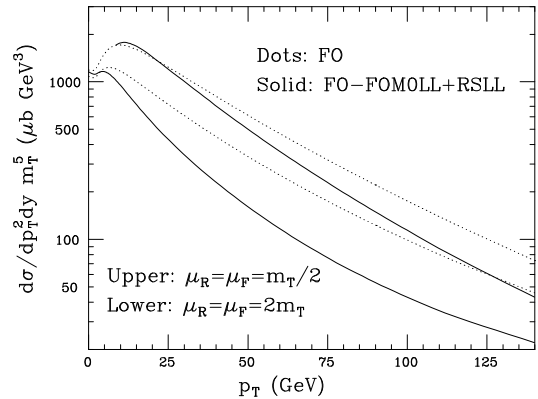}
    \caption{Matched caclculation at the leading logarithmic level.
             No improvement in the scale dependence can be observed.}
    \label{fig:band-ll-c0-mt5}
  \end{center}
\end{figure}

Our final result is given by the formula
{\equationcolor \begin{displaymath}
  \mbox{FONLL}=\mbox{FO}+(\mbox{RS}-\mbox{FOM0})\frac{\pt^2}{\pt^2+(c m)^2}.
\end{displaymath}}
This has the correct properties of being NLL accurate, and to include
mass effects up to the order $\as^3$. RS and FOM0
calculation obtained at the same transverse mass as the FO,
instead of same $\pt$ (for FO this is different)
to avoid large mismatch due to parton luminosity differences.\\
The value of $c$ that we pick has the following meaning: it is the
value of $\pt/m$ at which we believe that the massless formulae
make any sense at all. From fig.~\ref{fig:e_plot} it is apparent
that a value $c\approx 5$ is adequate.
Our final results are shown in fig.~\ref{fig:band-ms-c5}.
A slight anhancement for intermediate values of the transverse
momentum is present. At very large transverse momenta, a sensible
reduction of scale dependence is found.
At LL, we do not expect any improvement in scale dependence.
This is illustrated in fig.~\ref{fig:band-ll-c0-mt5}, where the resummed
calculation is included only at the LL level\footnote{Our result in this case
is at variance with the result of ref.~\cite{Scalise}.}.
\section{Conclusions}
The study of all order resummation of enhanced effects helps to
increase the precision of theoretical predictions in heavy flavour production.
NLL resummation of soft gluon effects reduces the theoretical
uncertainties in top cross sections for the Tevatron, and for
$b$ cross sections at HERAb.

Appropriately matched NLL resummation of large $\pt$ logarithms gives a
hint of slight increase in the predicted cross section,
going towards a better agreement with data.
Other ingredients of the calculation also go in the right direction.
Non-perturbative effects in fragmentation functions (the {\red
$\epsilon$ parameter} in the {\red Peterson Fragmentation function})
turn out to be smaller than previously thought. This is shown by the
LEP measurements of the $b$ fragmentation function, and by detailed
study of charm production in $\epem$ annihilation \cite{Cacciari97}.
A detailed study of the $\pt$ spectrum, with all modern ingredients,
is needed now.

\end{document}